\documentstyle[11pt]{article}
\input epsf

\newlength{\awidth}
\newlength{\aheight}
\newlength{\uswidth}
\newlength{\usheight}
\def\preprint#1{\gdef\@preprint{#1}}
\def\bce{\begin{center}}
\def\ece{\end{center}}
\def\be{\begin{equation}}
\def\ee{\end{equation}}
\def\bea{\begin{eqnarray}}
\def\eea{\end{eqnarray}}

\newcounter{fignr}
\newenvironment{fig}[1]{\refstepcounter{fignr}\label{#1}\begin{center}}{
    \end{center}}
\newcommand{\figcap}[2]{\parbox{#1}{{\footnotesize  Fig. \thefignr. #2}}}

\begin{document}
\baselineskip=.285in

%%%%%%%%%%%%%%%%%%%%%%%%%%%%%%%%%%%%%%%%%%%%%%%%%%%%%%%%%%%%%%%%%%%%%%%%%%%%%%
% title page
\catcode`\@=11
\def\maketitle{\par
 \begingroup
 \def\thefootnote{\fnsymbol{footnote}}
 \def\@makefnmark{\hbox
 to 0pt{$^{\@thefnmark}$\hss}}
 \if@twocolumn
 \twocolumn[\@maketitle]
 \else \newpage
 \global\@topnum\z@ \@maketitle \fi\thispagestyle{empty}\@thanks
 \endgroup
 \setcounter{footnote}{0}
 \let\maketitle\relax
 \let\@maketitle\relax
 \gdef\@thanks{}\gdef\@author{}\gdef\@title{}\let\thanks\relax}
\def\@maketitle{\newpage
 \null
 \hbox to\textwidth{\hfil\hbox{\begin{tabular}{r}\@preprint\end{tabular}}}
 \vskip 2em \begin{center}
 {\Large\bf \@title \par} \vskip 1.5em {\normalsize \lineskip .5em
\begin{tabular}[t]{c}\@author
 \end{tabular}\par}
 %\vskip 1em {\large \@date}
 \end{center}
 \par
 \vskip 1.0em}
\def\preprint#1{\gdef\@preprint{#1}}
\def\abstract{\if@twocolumn
\section*{Abstract}
\else \normalsize
\begin{center}
{\large\bf Abstract\vspace{-.5em}\vspace{0pt}}
\end{center}
\quotation
\fi}
\def\endabstract{\if@twocolumn\else\endquotation\fi}
\catcode`\@=12
%%%%%%%%%%%%%%%%%%%%%%%%%%%%%%%%%%%%%%%%%%%%%%%%%%%%%%%%%%%%%%%%%%%%%%%%%%%%%%

%\preprint{DPNU-94-55\\[0.5mm] SNUTP-96-031\\[0.5mm] WU-AP/57/96}
\preprint{}
\title{\Large\bf An Isolated Gravitational Dipole Moment Placed at The 
Center of the Two Mass Pole Model Universe
\protect\\[1mm]\  }
\author{\normalsize Eue Jin Jeong\\[1mm]
{\normalsize\it Department of Physics, The University of Texas at Austin, 
Austin, TX 78712}}

\maketitle

\def\gatij{\gamma^{ij}}
\def\gabij{\gamma_{ij}}
\def\ophi{\phi^{a}}
\def\hphi{\overline{\phi}^{a}}
\def\dint{\int\!\!\!\!\!\int}
\begin{center}
{\large\bf Abstract}\\[3mm]
\end{center}
\indent\indent
\baselineskip=.285in

The unexpected dynamic shift of the center of mass for a rotating 
hemisphere is shown to produce the general relativistic dipole 
field in the macroscopic scale.  This prompts us the question 
of what might be its cosmological implications.  The uniformly 
rotating sphere has the effect of the latitude dependent mass 
density distribution as reported by Bass and Pirani which is 
the cause of the `induced centrifugal force' in the Thirring's 
geodesic equation near the center of the rotating spherical 
mass shell.  On the other hand, one would expect the constant 
acceleration of the mass components may cause a general 
relativistic gravitational field.  The component-wise accumulation 
of this effect has been shown to appear as the non zero 
gravitational dipole moment in a rotating hemispherical mass shell. 
The present report discusses this non-Newtonian force 
experienced by a gravitational dipole moment placed at 
the center of the two mass pole model universe and its 
relevance to the observed anomalous red shift from far away galaxies.
\noindent
\newpage
\baselineskip=15pt
\pagenumbering{arabic}
\thispagestyle{plain}
\setcounter{section}{1}
\indent\indent
In 1911, von Laue [1] presented an argument of a general nature 
against the existence of a rigid body in relativity theory.  
It is based on the fact that a rigid body is expected to have 
a finite number of degrees of freedom, while, on the other 
hand, one can set up N disturbances near N separated points 
of the body, and they will be non-interacting, ie., 
independent for a sufficiently short interval of time 
(because of the finite propagation time required by relativity theory) 
so that there are at least N degrees of freedom, where N can be 
increased indefinitely for a continuous medium. 
	Other difficulties include the apparent paradox pointed 
	out by Ehrenfest [2].  Since the elements of the 
	circumference of a circle in a rotating disk move 
	in the direction of their instantaneous velocities, 
	one would anticipate a diminished value for the 
	circumference.  However, since the elements of 
	the radii of the circle everywhere move normally 
	to the direction of their velocities, no such 
	contraction for the length of the radius would 
	be expected.  This is the standard argument from 
	which it is concluded that the `geometry' of the 
	rotating body can not be Euclidean [3].  
	On the other hand, Hill[4] pointed out (as von 
	Laue [5] had done earlier) that a rotating body 
	would have a limit to the radius it could have, 
	for the velocity would vary linearly with the 
	radius and would exceed that of light for $r>1/\omega$ 
	and that the speed distance law must be nonlinear 
	while the Euclidean geometry is maintained. 
	However, Rosen [6] argued using a covariant 
	formulation that the speed distance law $(v=r\omega)$ 
	must be preserved and that the spatial geometry 
	on the surface of a rotating disk is non-Euclidean.  
	For these reasons, as pointed out by Phipps [7], 
	the rigidity of a metric standard or of any 
	extended structure such as a disk has never 
	found a consistent [8] relativistic definition 
	as a purely kinematic attribute. Rigidity has 
	therefore generally been conceived [9] as a 
	nonexistent physical property.  However, the 
	experiment suggested by Weinstein [10] and 
	later performed by Phipps [7] exhibited a 
	`rigid body' (to order $v^2/c^2$) and thus provides 
	evidence that rigidity is not always a physically 
	impermissible idealization.  He showed that 
	straight lines on the disk surface remain straight 
	on hyperplanes of constant laboratory time is 
	consistent with both the classical and the 
	Born[11] definition of rigidity.  
	The problem of the non-Newtonian gravitational 
	force experienced by a test particle inside a 
	rotating spherical shell has been considered by 
	Thirring [12] in 1918. In his calculation within 
	the weak field approximation, Thirring used the 
	constant mass density r and the four velocity 
	\\
\bea
u^1 &=& \frac{\omega R sin{\theta} sin{\phi}}{\sqrt{1 - \omega^2 R^2 
sin^{2}{\theta}}} \nonumber \\[5pt]
u^2 &=& \frac{\omega R sin{\theta} cos{\phi}}{\sqrt{1 - \omega^2 R^2 
sin^2{\theta}}} \nonumber \\[5pt]
u^3 &=& 0 \nonumber \\[5pt]
u^0 &=& \frac{1}{\sqrt{1 - \omega^2R^2 
sin^2{\theta}}} 
\eea
\\
and the length contraction 
\\
\be
d^3x' =d^3x''\sqrt {1 - \omega^2R^2 sin^2\theta}
\ee
\\
for the rotating spherical mass shell to perform the 
integration in the rest frame of the source to evaluate $\Phi_\mu^\nu$,
\\
\be
\Phi_\mu^\nu=4\int \frac{\rho u'_\mu u'^\nu d^3x'}
{ | \mathbf{r} - \mathbf{r}' | }
\ee
\\
from which $h_{\mu\nu}$ can be calculated as
\\
\be
h_{\mu\nu} = 
\Phi_{\mu\nu} - \frac{1}{2}\Phi
\ee
\\
In fact, one could have calculated the $\Phi_\mu^\nu$ in the rest 
frame of the observer by using the relativistic mass 
density in the same range of the radial integral and 
the resulting effects would have been the same since 
the integrand for $\Phi_\mu^\nu$ is the same for both cases.  By such 
a method of calculation[12], Thirring has effectively 
circumvented the problem of the questionable rigidity 
of the spherical mass shell.  On the other hand, 
physically, it is equivalent of taking the relativistic 
total mass-energy density $\gamma$($\omega,\theta$)$\rho$  for the dynamic mass 
components of the shell and then perform the integration 
in the observer's rest frame without concerning about 
the rigidity of the source.  

	For a test particle located close to the center 
	of mass of the rotating spherical mass shell of 
	radius R with the angular frequency $\omega$, the Cartesian 
	components of the acceleration (force/mass) have been 
	shown to be given by, using the above method [13] [14] [15],
\\
\bea
\ddot{x}&=&\frac{M}{3R}(\frac{4}{5}\omega^2 x - 8\omega\nu_y) 
 \nonumber \\[5pt]
\ddot{y}&=&\frac{M}{3R}(\frac{4}{5}\omega^2 y + 8\omega\nu_x) 
 \nonumber \\[5pt]
\ddot{z}&=&-\frac{8M}{15R}\omega^2 z 
\eea
\\
where $v_{y}$ and $v_{x}$ are the y and x components of the velocity 
of the test particle, respectively.  The test object placed 
slightly off the center of mass toward the z axis will be 
subjected to the harmonic oscillation according to the 
third formula given above.  The motion of the particle 
in the x, y plane is spiraling away from the rotation 
axis with the force proportional to $\omega^2r$ where r is the 
radial distance from the symmetry axis in the x, y plane, 
which has been interpreted as the indication of the 
existence of the induced centrifugal force in general 
relativity in accordance with the Mach's principle.   

	On the other hand, the field outside of the rotating 
	spherical mass shell has not been scrutinized in the 
	same fashion as Thirring did in his work on the 
	interior solution.  Part of the conceptual difficulty 
	was that it has been worked out in the general 
	formalism in the multipole expansion of the 
	linearized theory for $r>r'$ in which the Newtonian 
	potential has been derived and the conclusion has 
	been reached that the rotating spherical mass does 
	not have the dipolar effect [16] while Thirring's 
	work on the problem inside the rotating shell shows 
	interesting longitudinal linear force effect in the 
	z direction which is characteristically dipolar.  

	To elucidate the problem, it is necessary to reexamine 
	the dipole term in the weak field approximation for 
	the rotating hemispherical mass shell.  By following 
	the method of Thirring [12] outside of the source 
	$(r>r')$, assuming also that the mechanical stress 
	in the shell is small, it can be shown that the 
	dipole field calculated from the rotating hemispherical 
	shell has the non-zero value which cannot be 
	eliminated by the coordinate translation. By 
	keeping only the $T^{00}$, the corresponding gravitational 
	dipole moment is given by, in exact form
\\
\be
d_z= 
M\delta r_c=MR \left( \frac{1}{2} -
\frac {\frac {1 - \sqrt{1 - \alpha}}{\alpha}}{\sqrt{\frac {1}{\alpha}}
\sinh^{-1}{\sqrt{\frac{\alpha}{1 - \alpha}}}} \right)
\ee
\\
\be
\alpha= 
\frac{\omega^2 R^2}{c^2}
\ee
\\
for a bowl shaped hemispherical shell of radius R and 
mass M placed on the x, y plane, where $d_{z}$ is defined as 
the anomalous shift of the center of mass which doesn't 
depend on the choice of the specific coordinate system.  
For $\omega R<<c$, the $d_z$ can be approximated to be
\\
\be
\delta r_{c} = \frac{\omega^2 R^3}{24c^2}
\ee
\\

	As the result of this non zero gravitational 
	dipole moment, the field outside of the rotating 
	hemispherical mass shell of radius R is given by, 
	up to the approximation,
\\
\be
\phi= 
-\frac{M}{r}-\frac{d_z}{r^2} cos\theta + O(\frac{1}{r^3}) 
\ee \\	
where M is the total mass of the source, q is measured 
with respect to the positive z axis and $d_{z}$ is given by 
Eq. (6).  This dipole field has the force line which 
is exactly the same as that of the electric or magnetic 
dipole moment and diverges at $r=0$ since the expansion 
for $1/|r-r'|$ has been made with the assumption $r>r'$.  
The Cartesian components of the dipole force can be written by 
\\
\bea
F_{xdipole}&=&-\frac{\partial}{\partial x} (\frac{Gd_{z}}{r^2}\cos{\theta}) = 
Gd_{z}\frac{3zx}{r^5}
 \nonumber \\[5pt]
F_{ydipole}&=&-\frac{\partial}{\partial y} (\frac{Gd_{z}}{r^2}\cos{\theta}) = 
Gd_{z}\frac{3zy}{r^5}
 \nonumber \\[5pt]
F_{zdipole}&=&-\frac{\partial}{\partial z} (\frac{Gd_{z}}{r^2}\cos{\theta}) = 
Gd_{z}\frac{-{r^3} + 3{z^2}r}{r^6}
\eea
\\
where the center of the coordinate system is located at 
the distance $R/2$ from the center of the  sphere.  
The gravitational dipole moment points to the negative 
z axis (opposite to the direction of the center of mass 
shift) when the hemisphere is placed like a bowl on the 
x, y plane, independent of its direction of the angular 
velocity.  The direction of the z component of the force 
along the symmetry axis in Eq. (10) is uniformly toward 
the positive z axis except between the singular points.  
Thus, a test particle in front of the domed side of the 
hemispherical shell would be attracted toward inside and 
the one near the flat side will be repelled.  

	Using this result, one may attempt to calculate the 
	force experienced by a test particle at the position 
	(x, y, z), where $|x|=|y|=|z|<<R$, by attaching the two 
	rotating hemispherical shells to form a sphere and 
	adding the two opposite dipole forces near the center.  
	The center of the sphere is now at the boundary of the 
	hemisphere, separated uniformly from the shell by the 
	distance R.  Therefore, since the point of interest 
	is not exactly on the singular point of the volume 
	integral of the potential for the hemispherical 
	shell, one would expect the field near the center 
	of the sphere will behave reasonably well within 
	certain amount of expected error.  
	
	Since the dipole field is expressed in the coordinate 
	system in which the origin is located at the center 
	of mass of the hemisphere ($R/2$ from the center of 
	the sphere), the position of the test particle 
	becomes $(x, y, R/2-z)$ with respect to the coordinate 
	system the origin of which is located at the center 
	of mass of the upper hemisphere and $(x, y, R/2+z)$ 
	with respect to that of the lower one.  Using the 
	above equations and the relation (8) for the 
	expression for $d_{z}$, one obtains
\\
\bea
\ddot{x}&=&\frac{2M}{R}\omega^2 x 
\nonumber \\[5pt]
\ddot{y}&=&\frac{2M}{R}\omega^2 y 
\nonumber \\[5pt]
\ddot{z}&=&-\frac{4M}{R}\omega^2 z 
\eea
\\
for the force experienced by the test particle near 
the center of the rotating spherical mass shell of mass 
M for $\omega R<<c$.  
	Apart from the apparent formal resemblances, there 
	are couple of discrepancies between this result and 
	that of ThirringÕs.  The first conspicuous one is the 
	difference in the constant factor of 2/15 between the 
	two expressions. Also the information on the velocity 
	dependent force is lost which is caused by the fact 
	that the other components of the stress-energy 
	tensor have been ignored except $T^{00}$ for the field 
	outside of the source.  The discrepancy in the constant 
	factor would have been expected since the position 
	$r=R/2$ is not far outside of the boundary of the 
	source, while the $1/|r-r'|$ expansion for the dipole 
	moment was made with the assumption $r>r'=R/2$.  
	This problem is very similar to that of electromagnetic 
	vector potential from a circular ring of radius 
	a with current I.  It is well known that the 
	azimuthal component (the only non-zero term 
	due to the symmetry) of the vector potential 
	for both inside and outside of the radius a 
	of the ring is approximately given by 
\\
\be
A_\phi(r,\theta)=\frac{\pi I a^2 r sin \theta}{c(a^2 + r^2)^{3/2}}
\left( 1 + \frac{15 a^2 r^2 sin^2 \theta}{8(a^2 + r^2)^2} + ...\right)
\ee
\\

For $r>>a$, the leading term of this potential 
depends on $1/r^2$ indicating the dipole effect.  
It also gives the details of the potential inside 
the radius a without singularity.  Following this 
example, one may introduce a weight parameter $\eta$ 
into the gravitational dipole potential 
\\
\be
\phi_{dipole}\propto\frac{-r}{(\eta^2 + r^2)^{3/2}} 
\ee
\\
so that the potential behaves without singularity 
for $r<r'$, where $\eta$ represents the parametrized radius 
of the physical object.  For a non-spherical body 
like a hemisphere, for example, one may assign the 
parameter tentatively a virtual physical dimension 
of a shell 
\\
\be
\eta=\sqrt{0.1} R
\ee
\\
which is about one third of the radius R of the sphere.  
In this case, the corrected non-Newtonian force near 
the center of the sphere is reduced approximately by 
a factor 1/8 from the one in Eq. (11). This is close 
to the value 2/15 which gives exactly Thirring's induced 
centrifugal force. The above discussions suggest that 
the $\omega^2$ dependent forces in Thirring's result are mainly 
from the partially canceled dipole effect which arise 
due to the subtractive contribution to $F_{z}$ and the 
additive ones for $F_{x}, F_{y}$ from the two dipole moments 
respectively. 

	In regard to this problem, Bass and Pirani [17] 
	also have shown that the centrifugal force term 
	arises as a consequence of the latitude dependent 
	velocity distribution which generates an axially 
	symmetric (non-spherical) mass distribution, 
	which casts doubts on the centrifugal force 
	interpretation of the Thirring's result since 
	the rotating cylindrical object would not have 
	such latitude dependent density distribution and 
	there will be no corresponding centrifugal force 
	for the cylindrical object, contrary to our expectation.  
	These difficulties remain even when the contribution 
	from elastic stress is included, which led Bass and 
	Pirani to conclude that there was an apparent 
	conflict with Mach's principle.  Following this 
	observation, Cohen and Sarill reported that the 
	centrifugal term from Thirring's solution for a 
	rotating spherical mass actually represents a 
	quadrupole effect[18] by a deductive argument 
	and suggested an alternative solution [19] 
	(also previously by Pietronero[20]) for the 
	centrifugal force in general relativity using 
	the flat space metric in rotating coordinates. 

	The discussion so far indicates that the 
	superposed dipole field description gives 
	the quadrupole effect as proposed by Cohen 
	and Sarill which is identical to that of 
	Thirring's `induced centrifugal force' near 
	the center of the sphere. This also provides 
	a detailed look at the general field 
	configuration outside the rotating spherical 
	mass shell, up to the component $T^{00}$, which 
	is obtained by adding the fields created by 
	the two opposite dipole moments separated 
	by the distance R in addition to the monopole 
	field, 
\\
\be
\phi= 
-\frac{M}{r}+\frac{d_z/2}{|-(R/2)\hat{z}-\textbf{r}|^2}cos\theta'-
\frac{d_z/2}{|(R/2)\hat{z}-\mathbf{r}|^2}cos\theta''+O(\frac{1}{r^3})
\ee
\\
where the angles $\theta'$ and $\theta''$ are given by
\\ 
\bea
\theta'&=& tan^{-1}\left( \frac{r sin\theta}{r cos\theta + R/2}\right)
\nonumber \\[5pt]
\theta''&=&tan^{-1}\left(\frac{r sin\theta}{r cos\theta - R/2}\right)
\eea
\\
respectively and $d_z$ is given by the Eq. (6).  
By employing the result in Eq. (13), one may 
write the potential for a rotating spherical 
mass shell for both inside and out,
\\
\be
\phi= 
V(r)+\frac{|-(R/2)\hat{z}-\textbf{r}|d_z/2}{\lgroup\eta^2+(-(R/2)
\hat{z}-\textbf{r})^2\rgroup^{3/2}}cos\theta'-\frac{|(R/2)\hat{z}-\textbf{r}|d_z/2}
{\lgroup\eta^2+((R/2)
\hat{z}-\textbf{r})^2\rgroup^{3/2}}cos\theta''+O(\frac{1}{r^3})
\ee
\\
where 
\\
\be
\begin{array}{lcr}
\begin{array}{lll}
 V(r) & = & -M/r \\
 & = & -M/R \\
\end{array} &
\mbox{\hspace{1cm}}&
\begin{array}{l}
 \textnormal{for $r>R$} \\
 \textnormal{for $r\leq R$} \\
\end{array}
\end{array}
\ee
\\
and the angles $\theta'$ and $\theta''$ are given by  Eq. (16).  
	\\
% Fig 1:
\noindent\parbox{\textwidth}{\noindent\begin{fig}{fig}
  \mbox{\setlength{\epsfxsize}{.80\textwidth} \epsfbox{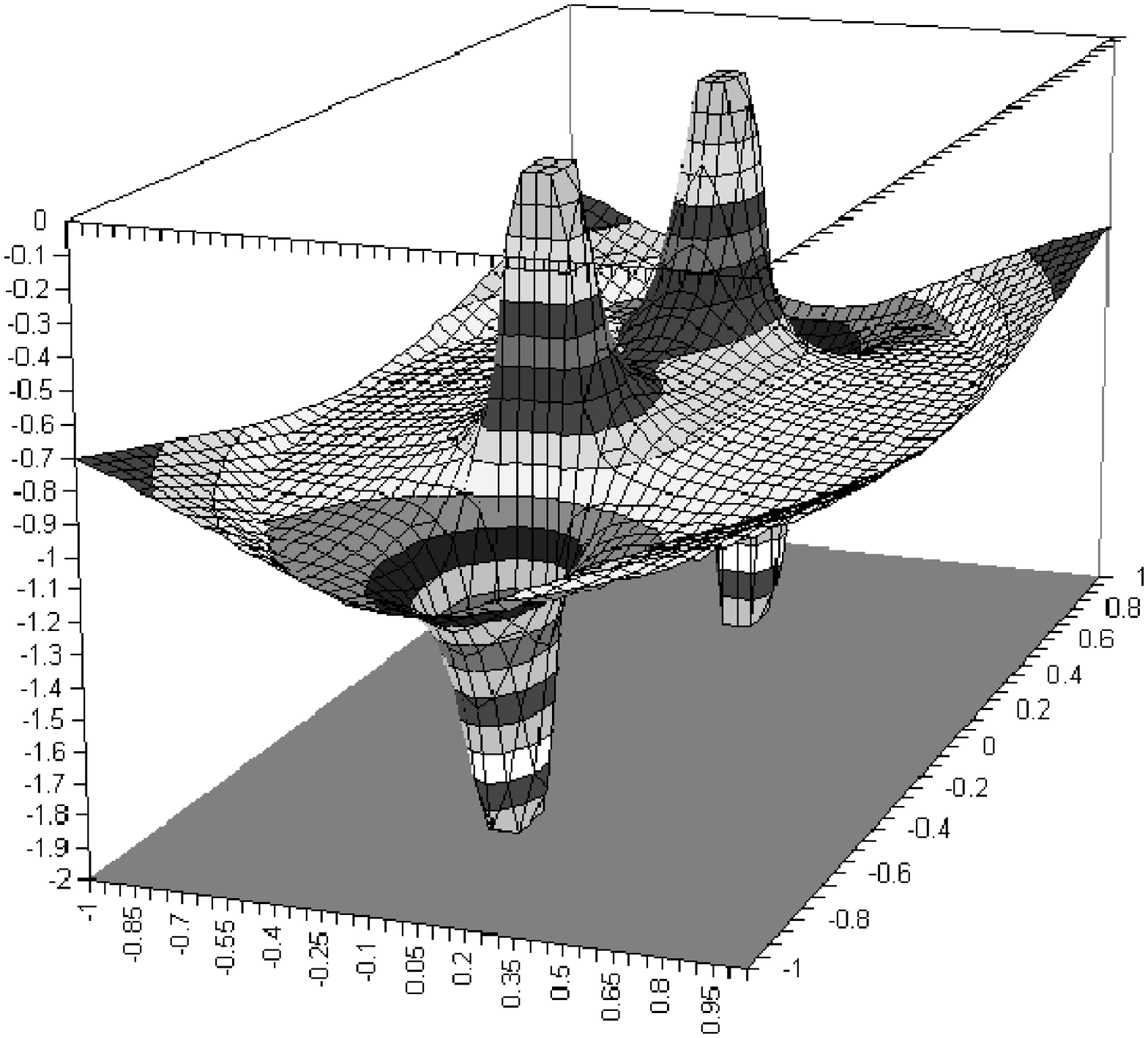} }\\
\ \\
  \figcap{.9\textwidth}{A close up view of the computer simulated 3-D diagram 
for the multipole gravitational potential inside a spherical 
mass shell rotating along the z axis.  All the constants are 
set equal to 1 ($M=G=R=c=1$) and the value of the parameter $\eta$ 
is set equal to zero. The anomalous center of mass shift $\delta r_c$
for the potential in the diagram is 0.05R which corresponds 
to the case $\omega R=0.045 c$. The maximum variable range of $\delta r_c$ 
is from 0 to 0.5R.}
\end{fig}
}\\ 

	The potential diagram in Figure shows four distinctive poles 
	indicating that it is indeed a quadrupole 
	moment created by superposing two opposite dipoles.  
	One of the major consequences of this potential 
	is that there exist two strong repulsive 
	potential peaks close to the center along 
	the z axis, which is unusual for the gravitational 
	force.  Since it is generally believed that the 
	repulsive gravitational potential can not be 
	generated by ordinary matter, this is a puzzling 
	event since there exists a pair of virtual 
	negative mass poles as well as the positive 
	ones inside a rotating spherical mass shell.  
	The saddle point created by this two positive 
	peak potentials near the center is the cause of 
	ThirringÕs `centrifugal force' as discussed above.  
	This quadrupole feature disappears as soon as the 
	rotational frequency becomes zero as can be seen 
	from Eq. (17). 
	
    The behavior of the potential in the longitudinal 
	axis also indicates the possibility that the particles 
	traveling into the attractive dipole potential 
	well along the z axis will be repelled back to 
	where they have come from depending on the rotational 
	frequencies that support the height of the peaks. 
	This repulsive potential peak determines the range 
	of the linear orbital distances that the particles 
	may travel back and forth from the poles to the far 
	outsides along the z axis.  
	
	Since there is no 
	compelling evidence that the plasma and magnetic 
	field must be generated inside rotating ultra-compact 
	bodies, where the electronic orbital states have 
	been long before collapsed, one may suspect that 
	the superposed dipole effect may have been the 
	major driving force behind the jet phenomenon 
	in some of the fast rotating cosmological 
	objects.  This view point is supported by the 
	fact that the dipole field is long ranged and 
	the strongest next to that of the monopole.  
	The long range potential dip around the equator 
	in the diagram also indicates that there exists 
	a tendency of the cluster formation around the 
	equatorial plane of rotating celestial bodies.  

	The information that has been lost by neglecting 
	the component $T_{ik}$ includes the one from the 
	elastic stress which has been considered by 
	Bass and Pirani [17] as discussed above and 
	also the velocity dependent (Coriolis) force 
in Eq. (5) which depends linearly on the 
rotational frequency $\omega$. Contributions from 
these terms may be added into the potential 
without the loss of generality of the strong 
dipole feature.  On the other side of the 
problem, the effect of the `centrifugal 
force' from rotating spherical objects has 
been investigated by many researchers and 
a result has been recently reported by Gupta et 
al. [21].  As r increases in the equatorial 
plane, the saddle potential progresses 
toward the plateau which is a constant 
potential region.  As can be seen from 
Figure, this transition is smooth without 
abrupt change of curvature.  The force 
reaches the peak at a certain radius $r_i$ 
from the center where the slope of the 
potential is the steepest and then 
gradually becomes zero as the potential reaches 
the plateau.  This prediction of the general 
behavior of the maximum of the `centrifugal 
force' from the potential (16) and (17) is 
consistent with the result (Fig.1) of Gupta 
et al. [21].  The reversal (sign change) effect 
is not present in this case since the object is 
made of a hollow shell.  
	Although the main cause of the above result may seem like 
	due to the mass increase from
	the latitude dependent velocity distribution 
	of a rotating spherical mass as reported by 
	Bass and Pirani [17], the corresponding permanent increase of the 
	mass density
	would not produce the dipole effect since there would 
	not be the dynamic shift of the center of mass in such case.  
	Therefore, it is clear that this dynamic effect is due to 
	the collective contribution from the
	uniform acceleration of the mass components in the rotating sphere.
	
	The above discussion was based on the 
	fact that the analytic interior solution of 
	a rotating spherical mass shell exhibits 
	a multipole potential which suggests there exist 
	physically meaningful gravitational 
	dipole moment in our universe contrary 
	to the general belief that it doesn't.  
	This argument is supported by the observation 
	that the linear orbitals of the particles 
	along the two poles of a rotating spherical 
	mass resembles closely the observed jets from the fast 
	rotating cosmological bodies.  
	On the other hand, it becomes immediately 
	obvious that a rotating asymmetrical body 
	can have isolated gravitational dipole 
	moment contrary to the generally known 
	interpretation of the linearized theory.  
	An independent dipole moment posseses 
	all the dynamics that an isolated, 
	controllable magnet would behave in the 
	magnetic monopole universe. 

	In relation to this problem, one of the 
	widely known mysteries of our current 
	cosmology is the presence of the anomalous 
	red shift observed from some of the far 
	away galaxies.  This has been a serious 
	problem in cosmology since the current 
	model of the universe does not allow such 
	mode of movement of a galaxy assuming that 
	the relativistic Doppler shift correctly 
	represents the relative velocity between 
	the observer and the observed.  The theory 
	of the big bang associated with HubbleÕs 
	expansion law prohibits motions of the 
	galaxies other than the uniform separation 
	between any two of the galaxies.  
	The conceptual model of an isolated 
	gravitational dipole moment placed in 
	the matter filled universe actually solves 
	the problem of the anomalous red shift by 
	simply assuming that the specific galaxy 
	possesses the non zero net gravitational 
	dipole moment in the direction of our 
	galaxy.   If we assume that at the time 
	of the big bang some of the chunk of the 
	matter came off with an asymmetric shape 
	of the body with non zero rotational 
	frequency, they will eventually acquire 
	velocities unrelated to the uniform 
	Hubble expansion.  Even if the detached 
	body may change its shape in time to 
	become a longitudinally symmetric structure, 
	the accumulated linear momentum will remain 
	to be observed as an anomalous red shift 
	in the spectrum.  

	Assuming that the universe can be modeled 
	to consist of two large mass poles in the 
	front and back of the dipole separated by 
	the distance r with mass M/2 where M is 
	the mass of the universe, one can calculate 
	the net directional force on the dipole moment to be
\\
\be
F=\frac{2dM}{(r/2)^3} 
\ee
\\
where M is the mass of the universe and r the 
distance between the two mass poles.  The direction 
of the force is toward the positive (pointed) side 
of dipole.  The influence from the monopole 
force is canceled because the dipole moment is 
placed exactly in the middle of the two mass 
poles of the model universe.  The next question 
is if the observed galaxies which exhibit anomalous 
red shift indeed have asymmetrical shape with 
respect to its longitudinal axis.  It is something 
that may need to be verified by observational 
astronomy if indeed they do have asymmetrical 
configuration along the rotation axis.  
	From the above considerations, it is obvious 
	that we have a much more flexible model 
	of the universe by including the dipole 
	force in our system.  The nature of this 
	force is that, while it has been formally 
	predicted by general relativity, it has 
	not been fully recognized by the scientific 
	community because the traditional treatment 
	of the weak field approximation for a 
	spherically symmetric body has concluded 
	that there is no dipole term.  However, 
	since ThirringÕs solution for the `induced 
	centrifugal force' turned out to represent 
	the partially canceled dipole field inside 
	a rotating spherical mass shell, it no 
	longer justifies to neglect the gravitational 
	dipole moment.    
	
	The major consequences of this solution 
	is that an axisymmetric yet longitudinally 
	asymmetric rotating object can have the 
	isolated dipole gravitational moment which 
	is virtually identical in dynamics to 
	the magnetic dipole moment placed in a sea 
	of the uniform magnetic monopoles.  
	By this result, two of the main cosmological 
	mysteries of our time have become trivial 
	consequences of the gravitational dipole moment 
	that is imbedded in general relativity. 

\def\hebibliography#1{\begin{center}\subsection*{References}
\end{center}\list
  {[\arabic{enumi}]}{\settowidth\labelwidth{[#1]}
\leftmargin\labelwidth	  \advance\leftmargin\labelsep
    \usecounter{enumi}}
    \def\newblock{\hskip .11em plus .33em minus .07em}
    \sloppy\clubpenalty4000\widowpenalty4000
    \sfcode`\.=1000\relax}

\let\endhebibliography=\endlist

\begin{hebibliography}{100}

\bibitem{B1}M. von Laue, Relativitaetstheorie  (1921), fourth edition, Vol. {\bf 1}, pp.203-204
\bibitem{B2}P. Ehrenfest, Physik. Zeits., {\bf 10}, 918 (1909)
\bibitem{B3}A.Einstein and L. Infeld, The Evolution of Physics (Simon and Shuster, New York, 1938), p. 240
\bibitem{B4}E.L. Hill, Phys. Rev., {\bf 69}, 488 (1946)
\bibitem{B5}M. von Laue, Relativitaetstheorie  (1921), fourth edition, Vol. {\bf 1}, p. 24
\bibitem{B6}N. Rosen, Phys. Rev., {\bf 71}, 54 (1947) 
\bibitem{B7}T. E. Phipps jr., Lettere Al Nuovo Cimento, {\bf 9}, 467(1974)  
\bibitem{B8}G. Cavalleri, Nuovo Cimento, {\bf 53} A, 415 (1968)
\bibitem{B9}H. Arzelies, Relativistic Kinematics (New York, 1966)
\bibitem{B10}D. H. Weinstein, Nature, {\bf 232}, 548 (1971) 
\bibitem{B11}M. Born,  Ann. Phys., {\bf 30}, 840 (1909)
\bibitem{B12}H. Thirring, Z. Phys., {\bf 19}, 33 (1918); and {\bf 22}, 29 (1921)
\bibitem{B13}J. Weber, General Relativity and Gravitational Waves (Interscience Publisher, Inc., New York, 1961), p. 160
\bibitem{B14}L. D. Landau and E. Lifshitz, The Classical Theory of Fields (Addison-Wesley Publishing Company, Inc., Reading, Massachusetts, 1959)
\bibitem{B15}C. Moeller, The theory of Relativity (Oxford University Press, London, 1952), pp. 317 ff.
\bibitem{B16}C. W. Misner, K. S. Thorne and J. A. Wheeler, Gravitation (Freeman, San Francisco, 1973), p 991.
\bibitem{B17}Bass, L., and Pirani, F. A. E., Phil. Mag., {\bf 46}, 850 (1955)
\bibitem{B18}J. M. Cohen  and W.J. Sarill,  Nature, {\bf 228}, 849 (1970)
\bibitem{B19}J. M.Cohen, W. J. Sarill  and C. V. Vishveshwara, Nature, {\bf 298}, 829 (1982)
\bibitem{B20}L. Pietronero, Ann. Phys., {\bf 79}, 250(1973)
\bibitem{B21}Anshu Gupta, Sai Iyer and A. R. Prasana, Class. Quantum Grav., {\bf 13}, 2675 (1996); see also the references therein

\end{hebibliography}
\end{document}